\documentstyle[prb,preprint,aps]{revtex}

\begin{document}
\draft
\title{Electronic Theory for the Transition from \\
       Fermi-Liquid to Non-Fermi-Liquid Behavior \\
       in High-T$_{c}$ Superconductors}

\author{M. Langer, J. Schmalian, S. Grabowski, and
 K.H. Bennemann}
\address{Institut f\"ur Theoretische Physik,
  Freie Universit\"at Berlin, Arnimallee 14, \\
       14195 Berlin, Germany}
\date{August 15, 1995}
\maketitle
\begin{abstract}
  We analyze the breakdown of Fermi-liquid
  behavior within the 2D Hubbard model as function of doping
  using our recently developed numerical method for the
  self consistent summation of bubble and ladder diagrams.
  For larger doping concentrations the system
  behaves like a conventional Fermi-liquid and for
  intermediate doping similar to a marginal Fermi-liquid.
  However, for smaller doping pronounced
  deviations from both pictures occur which are due to
  the increasing importance of the short range
  antiferromagnetic spin fluctuations. This is closely related
  to the experimental observed shadow states in the normal
  state of
  high-$T_c$ superconductors.
  Furthermore, we discuss the implications of our results
  for transport experiments.
  \vskip 5mm
  \noindent
  Key words: A. High-T$_{c}$ superconductors
\end{abstract}

\newpage

Central questions in the theory of the high-$T_c$
superconductors are related to the Fermi-liquid (FL) properties.
Although the existence of a Fermi surface (FS) is well
established~\cite{OLL90}, pronounced deviations from
FL behavior could be observed.
For strongly overdoped compounds the experimental
results seem to be consistent with a standard FL
picture, whereas for systems near the optimal doping
this is clearly not the case:
indications for a linear energy-dependence of the inverse
quasi particle
lifetime reported in angle resolved photoemission (ARPES)
measurements~\cite{OLL90}, the unusual temperature
dependence of
transport coefficients~\cite{BHT94,HBT94} and the $\omega /T$
scaling behavior of the dynamical magnetic susceptibility
observed in neutron scattering measurements~\cite{KBB92}
can not be understood within the conventional FL
picture.  For underdoped systems the
deviations from FL behavior are even stronger
due to the dominance of short range antiferromagnetic
correlations.
To achieve a comprehensive understanding of the normal
state properties of the high-T$_{c}$  superconductors
the marginal Fermi-liquid (MFL) theory was
proposed~\cite{VLS89}.
In this phenomenological approach the
characteristic energy scale is given only by the temperature
and the inverse lifetime of the quasi particles is proportional
to their excitation energy. However, this ansatz does not take
into account the strong short range antiferromagnetic
correlations
in the $CuO_{2}$ planes,
whose influence on the quasi particle properties
was demonstrated in NMR~\cite{ISY93}, neutron
scattering~\cite{BGJ88}
and also in ARPES measurements~\cite{SD95,AOS94}.

Interesting insight
into the structure of these highly correlated
electron systems could be obtained from Monte Carlo
and exact
diagonalization studies~\cite{D94}.
An alternative approach to investigate the dynamical
properties
of the high-$T_{c}$ systems is offered by perturbation
theory.
To achieve an appropriate description of the
low energy spin fluctuations we
use in the following the fluctuation exchange
approximation (FLEX)
by Bickers and Scalapino~\cite{BS89,BSW89}.
This self consistent summation of all bubble and ladder
 diagrams
is a conserving approximation in the sense of Baym
and Kadanoff~\cite{BK61}.
In particular, it takes into account the interaction of the
quasi particles
with spin fluctuations which are expected to be the
 dominating low
energy excitations in the high-$T_c$ materials.
In distinction to other diagrammatic methods the self
consistency allows
to consider larger interaction strengths, whereas in
distinction to
exact diagonalization
calculations and quantum Monte Carlo simulations
we are able to use much larger lattices which gives access to
the low energy properties with negligible finite size effects.

Since we are interested in dynamical properties
(especially in the quasi particle lifetimes),
we transform the FLEX-equations, which are formulated
on the
imaginary frequency axis, to the real axis~\cite{SLG95}.
Thus we determine by calculating the
momentum dependent Greens function $G_{\bf k}(\omega)$
and
its self-energy $\Sigma_{\bf k}(\omega)$
the low energy properties within the framework
of the 2D Hubbard model.
This enables us to study the character of the elementary
excitations as a function of the doping concentration
$x=1-n$, where $n$ is the occupation number per site.
We present new results on (i) the continuous breakdown of
the Fermi-liquid picture for decreasing doping
(ii) the interplay between this phenomenon and the
occurrence
of shadows of the Fermi-surface and
(iii) the temperature scale set by the spin fluctuations
and its influence on the temperature dependence of the
Hall-coefficient.

We consider the one-band Hubbard Hamiltonian
\begin{equation}
H=\sum_{ij \sigma} (t_{ij}-\mu \delta_{ij})
   c^\dagger_{i \sigma}c_{j \sigma}
   +U \sum_i c^\dagger_{i \uparrow }c_{i \uparrow }
    c^\dagger_{i \downarrow }c_{i \downarrow } \,
\label{Hub}
\end{equation}
where $c^\dagger_{i \sigma}$ is the creation operator
of an electron
at lattice site $i$ and with spin $\sigma$.
$t_{ij}$ is the hopping matrix element between
sites $i$ and $j$,
$\mu$ is the chemical potential and $U$ is the local
Coulomb repulsion.
The results presented below are obtained for
nearest neighbor hopping
$t=0.25 \, {\rm eV}$ and $U=4t$.
The self-energy of this Hamiltonian within
 the FLEX-approximation,
neglecting the particle-particle excitations
which were shown to be of minor
importance~\cite{BSW89},
is given by the following momentum
and Matsubara-frequency  sum~\cite{BS89}
$
\Sigma_{{\bf  k} }(i\omega_n)=\frac{T}{  N}
\sum_{{\bf  k}',n'}
 G_{{\bf k}'}(i\omega_{n'})
V_{{\bf k}-{\bf k}'}(i\omega_n-i\omega_{n'}) \,
$
where $T$ is the temperature and $N$ the
number of sites of the
finite lattice.
The effective interaction $V_{{\bf k}}(i\nu_m)$
 resulting from the
summation of bubble and electron-hole ladder
 diagrams is given by
\begin{eqnarray}
V_{{\bf q} }(i\nu_m )=  \frac{U^2}{2} \chi^o_{{\bf q} }(i\nu_m )
\left(\frac{3}{1-U\chi^o_{{\bf q} }(i\nu_m )}
  +    \frac{1}{1+U\chi^o_{{\bf q} }(i\nu_m )} - 2 \right) \, .
\label{effint}
\end{eqnarray}
Here, $\chi^o_{{\bf q} }(i\nu_m )= -\frac{T}{  N} \sum_{{\bf  k},n}
 G_{{\bf k}+{\bf q}}(i\omega_{n}+i\nu_m)
  G_{{\bf k} }(i\omega_{n} ) $ is the particle-hole
bubble and
 $ \omega_n=(2n+1)\pi  T $ and  $ \nu_m=2m\pi  T $
are the fermionic
and bosonic Matsubara frequencies, respectively.
Furthermore, the Greens function is given by the
Dyson equation
$G_{{\bf  k} }(i\omega_n) = \left( i\omega_n +
\mu - \varepsilon({\bf k})
                         - \Sigma_{{\bf  k} }(i\omega_n)
 \right) ^{-1}$
where $\varepsilon({\bf k}) = -2t (\cos(k_{x}) +
 \cos(k_{y}))$
is the free dispersion.
These equations are analytically transformed
to the real frequency axis
yielding a set of equations for the Greens function
$G_{{\bf  k} }(\omega)$ and the self-energy
$\Sigma_{{\bf  k} }(\omega)$, which is solved
self consistently~\cite{SLG95}.

In Fig.1 we present our results for a system with
 doping $x=0.10$ at
$T=63 K$. Fig.1(a) shows ${\rm Im }\,
\Sigma_{\bf k}(\omega)$ for
momenta close to the FS (solid circles in the
 schematic plot of the
Brillouin zone in the inset of Fig.1c).
Note that $\omega = 0$ corresponds to the
Fermi energy.
For low energy hole ($\omega < 0$) as well as
for low energy particle ($\omega > 0$) excitations a linear
$\omega$-dependence for energies $|\omega| > \omega'$
where $\omega' \approx \, 8 \,{\rm meV}$ can be seen.
Note that $\omega'$ is of the order of the temperature
broadening
$T = 63 K \cong 6 \,{\rm meV}$.
Although this behavior is consistent with the MFL theory
the different slopes for hole excitations and for particle
excitations are in distinction to MFL.
In Fig.1(c) we plot the corresponding real part of the
self-energy and
in Fig.1(e) the spectral density, which consists only of a main
peak.

In Fig.1(b) we plot
${\rm Im }\, \Sigma_{\bf k}(\omega)$ for momenta ${\bf k}$
close to the
Fermi surface shadow (FSS), which are indicated as open
circles in the
inset of Fig.1(c).
The FSS consists of those points ${\bf k'}$ of the Brillouin
zone
which are shifted by ${\bf Q} =(\pi,\pi)$ relative to the FS.
In a recent paper~\cite{LSG95} we discussed the
formation of shadow
states resulting from short range antiferromagnetic
correlations.
Due to the strong coupling of quasi particle states
with momenta
${\bf k} + {\bf Q}$ at the FSS to those near
${\bf k}$ at the main FS spectral weight is transferred
 from
${\bf k}$ to ${\bf k} + {\bf Q}$ which results in the
 formation of
shadow states. We found this coupling to be mostly
 pronounced
for FS and FSS states nearest to the $(\pi,0)$-point,
resulting from the flat bands in this region.
At the FSS, the deformations of ${\rm Im }\,
\Sigma_{\bf k}(\omega)$
relative to MFL behavior are much
stronger compared to FS states and a double
 peak structure develops,
which is mostly pronounced for ${\bf k} = (\pi,\pi/8)$
(indicated as (II) in the inset of Fig.1(c)).
In Fig.1(d) we plot the real part of the self-energy
for the same
momenta on the FSS.
The local maxima close to the FS are clearly visible,
but not strong enough to produce new poles of the
Greens function,
which would result in new antiferromagnetically
induced
quasi particle states as proposed by Kampf
and Schrieffer \cite{KS90}.
Nevertheless, satellites in the density of states
occur
very close to the Fermi energy (Fig.1(f)),
which build up the experimentally observed
shadows
of the FS \cite{AOS94}.
For momenta far away from the FS as well as
from its shadow we
find the FL-like quadratic frequency dependence
of
${\rm Im }\, \Sigma_{\bf k}(\omega)$ (not shown).
{}From the ${\bf k}$-dependence of the spin
susceptibility we
calculate a rough estimation of the
correlation length $\xi$ by neglecting vertex corrections
\cite{Corr_Len}. For $x=0.10$ we find
$\xi \approx 3.5a_{0}$, where $a_{0}$ denotes
 the lattice spacing in
the $CuO_2$ planes, in agreement with the
experimental observations.

For larger doping, the correlation effects in
the high-$T_{c}$ compounds
are expected to get weaker resulting in
conventional FL
behavior. Therefore, we investigated the
doping dependence of the
self-energy.  In Fig.2 we show ${\rm Im }\,
\Sigma_{\bf k}(\omega)$
for a FS state and for a FSS state
for $x=0.10$ and $x=0.20$.
For the FS momentum (solid lines)
in both cases there exists a crossover
from a quadratic to a linear $\omega$-dependence.
However, for
$x=0.10$ the crossover takes place at
$\omega \approx 8 \,{\rm meV}$
(which is of the order of the temperature,
see discussion above),
whereas for $x=0.20$ it happens at the
much higher value $\omega \approx 25 \, {\rm meV}$.
For increasing doping this crossover
takes place at even higher energies
resulting in FL behavior for
$x > 0.25$ (not shown).
Now we discuss the frequency dependence at
the FSS (dashed lines).
For $x=0.20$ the spectral density $\rho_k(\omega)$
is small for $\omega$ close to the Fermi level
because shadow states arise only for low doping
concentrations.
Consequently, states shifted by ${\bf Q}$ relative
to the FS exhibit
the frequency dependence
${\rm Im }\, \Sigma_{\bf k}(\omega) \propto \omega^{2}$
like all other states which are far away from the FS
as can be seen in Fig.2(b).
However, for $x=0.10$ (Fig.2(a)) strong deviations
from this behavior
occur due to antiferromagnetic fluctuations,
as discussed in the preceding paragraph.

The new structures of the self-energy, especially
the double well
in ${\rm Im }\, \Sigma_{\bf k}(\omega)$ for momenta
close to the
FSS, show
that in contrast to the MFL theory the temperature
is not the only energy scale in this model. A second
characteristic energy scale is set by the spin fluctuations
themselves. To investigate the interplay of these
two energy scales in more detail we solved our equations for a
fixed doping $x=0.11$ and various temperatures.
In Fig.3(a) we show  ${\rm Im }\, \Sigma_{\bf k}(\omega)$ for
a momentum close to
the FS (solid circle labeled (I) in the inset of Fig.1(c)).
For increasing temperature the linear $\omega$-dependence
is replaced by a functional form like
${\rm Im }\, \Sigma_{\bf k}(\omega) \propto
 (\omega - \omega_{0})^{2}$
in an energy range of the order of $T$.
The different slopes of the inverse quasi particle lifetime
for electron respectively hole excitations are for higher
temperatures reflected in a shift $\omega_{0}$ of
the maximum
of ${\rm Im } \, \Sigma_{\bf k}(\omega)$
away from the Fermi energy.
Furthermore, an extrapolation to even lower temperatures
indicates that ${\rm Im }\, \Sigma_{\bf k}(\omega=0)$ as
function of $T$ does not saturate to $0$ for
$T \longrightarrow 0$ as in the FL picture but to a finite
value.
Such a behavior is consistent with our picture that the
shadow states
arise from quasi particle decay due to the strong
coupling of states
near to the FS to states near to its shadow. Of course,
this coupling is
also present at $T=0$.
In Fig.3(b) we show ${\rm Im }\, \Sigma_{\bf k}(\omega)$
for the same doping and
for a momentum close to the FSS
(open circle labeled (II) in the inset of Fig.1(c)).
The double well structure vanishes at $T \approx 300 K$.
Note that ${\rm Im }\, \Sigma_{\bf k}(\omega=0)$ does
not only deviate from
conventional FL or MFL behavior,
but even has a maximum for $T \approx 300 K$
and afterwards decreases for decreasing temperature,
again
yielding for $T \longrightarrow 0$ a finite value.
For states far away from the FS and from its shadow
we detected a quadratic temperature dependence
(not shown).
The existence of a new energy scale can also be seen
in the momentum averaged density of states (not shown).
Here, the spin density wave pseudogap, which is clearly
visible for
low temperatures ($T=75 K$), becomes smaller with
increasing
temperatures and vanishes at $T \approx 750 K$.
Therefore, we conclude that this temperature represents
for $x=0.11$ the characteristic energy of the spin fluctuations
\cite{Finite_Size}.

The unusual momentum, frequency and temperature
dependence of the elementary
excitations has important consequences for the
transport properties
of the high-$T_{c}$ superconductors. The linear
 $\omega$-dependence
of ${\rm Im }\, \Sigma_{\bf k}(\omega)$ for
the FS states (see discussion of Fig.1(a)) corresponds to
a linear temperature dependence of the electrical
resistivity~\cite{WT93}
for materials near the optimal doping concentration
in agreement with
the experiments.
However, for smaller doping we found deviations from
the linearity of ${\rm Im }\, \Sigma_{\bf k}(\omega)$.
Moreover, shadow states with very unconventional
properties
of the scattering rate get occupied that can
additionally influence the transport coefficients.
This results in deviations from the linear temperature
behavior of the
electrical resistivity.
Furthermore, due to the coexistence of different
temperature
dependences for different scattering modes the Hall
coefficient becomes $T$-dependent \cite{A91,CML92}.
To demonstrate that our self consistent FLEX-calculation
is able to describe such an effect we plot in the inset of Fig.3(b)
our results for the Hall coefficient obtained by neglecting
further vertex corrections for the conductivity within the
magnetic field \cite{NH76} for doping
$x=0.11$ and various temperatures.
$R_{H}(T)$ decreases for increasing
temperature and saturates above the characteristic
spin fluctuation
energy of $T \approx 750 K$ in agreement with the
experiments~\cite{HBT94}.

In conclusion, we investigated the properties of the
 elementary
excitations in the normal state of high-$T_{c}$
superconductors in the framework of
FLEX and the one-band Hubbard model.
Due to our new numerical method we calculated
the self-energy directly
on the real frequency axis and obtained new results for
the fine structure of this quantity as function of doping.
We found its frequency and temperature behavior
to be strongly momentum dependent and to exhibit
very unconventional structures especially for states at the
Fermi surface shadow, which get more pronounced for
 lower doping.
We obtained a continuous crossover from
Fermi-liquid behavior for
very large doping ($x > 0.25$) to an intermediate range
which behaves similar to a marginal Fermi-liquid
($x \approx 0.15$)
and finally to a doping range
where the system is dominated by strong antiferromagnetic
correlations ($x \approx 0.1$).
For increasing temperature we find in all doping
regions a transition to Fermi-liquid behavior.
The doping and temperature dependence of the Fermi-liquid
properties is schematically illustrated in Fig.4.
Note, that the phase diagram of our theory
is in agreement with the normal state phase diagram
which is obtained by analyzing the transport properties of
$(La,Sr)_2 Cu O_4$. \cite{BHT94}
In a forthcoming paper \cite{GSL95} we investigate the
 consequences of this
qualitatively new behavior of the quasi particle states
for superconductivity.
%
%
%
%

\newpage
\begin{figure}
\caption{Self-energy and spectral density for $x=0.10$ and
         $T = 63 K$.
     (a) ${\rm Im }\, \Sigma_{\bf k}(\omega)$ for four
     different momenta which are
     close to the Fermi-surface. Note the linearity up
     to very small frequencies.
     (b) The same quantity for momenta close to the
     Fermi surface shadow.
     The double well structure is mostly pronounced for ${\bf k}$
     nearest to the $(\pi,0)$-point.
     (c) ${\rm Re}\, \Sigma_{\bf k}(\omega)$ for momenta close
     to the Fermi surface.
     (d) The same quantity for momenta close to the
     Fermi surface shadow.
     (e) Spectral density along the main Fermi surface.
     (f) Spectral density along the Fermi surface shadow.
     Inset of (c) Brillouin zone for orientation. Momenta on the
     main Fermi surface are indicated by solid circles, those
     on the shadow Fermi surface by open circles.}
\end{figure}

\begin{figure}
\caption{Imaginary part of the self-energy for fixed
     temperature
      $T=63 K$ and two doping concentrations.
     (a) ${\rm Im }\, \Sigma_{\bf k}(\omega)$ for small
     doping ($x=0.10$).
     The solid line indicates a momentum on the main
     Fermi surface
     (solid circle labeled (I) in the inset of Fig.1(c)),
     the dashed line a momentum on its shadow
     (open circle labeled (II) in the inset of Fig.1(c)).
     Note the strong deformations relative to Fermi-liquid
     and marginal Fermi-liquid behavior at the shadow.
     (b) ${\rm Im }\, \Sigma_{\bf k}(\omega)$ for larger
     doping ($x=0.20$)
     with solid and dashed line as in (a).
     The frequency dependence
     of both is Fermi-liquid like.
         }

\end{figure}

\begin{figure}
\caption{Imaginary part of the self-energy or $x=0.11$
         and different temperatures.
         In (a) we plotted ${\rm Im }\, \Sigma_{\bf k}(\omega)$
          for a momentum close
         to the Fermi surface, in (b) for a momentum close
         to the Fermi surface
         shadow. The double well structure vanishes
         $T \approx 300 K$.
         Note that the scattering rates are large for both
         cases and for all temperatures we considered.
         In the inset of (b) we show the Hall-coefficient
         for the same
         doping as a function of $T$.}
\end{figure}

\begin{figure}
\caption{Schematic diagram for the transition from
         Fermi-liquid
         to non Fermi-liquid behavior in the normal states
         for intermediate coupling $U=4t$ as discussed in
         the text.
         The boundaries of the various areas are diffuse
         and there
         is a continuous crossover between the three regions.
         The smallest doping for which our calculations
         were done is $x=0.09$. Therefore, the
         antiferromagnetic phase transition observed in
         the cuprates for even lower doping is not included
         in the diagram.
         Note that our results are consistent with the
         Mermin-Wagner theorem~\protect\cite{MW66}
          because we found no
         antiferromagnetic instability.}
\end{figure}


\begin{thebibliography}{9}
\bibitem{SLG95}
  Details will be published elsewhere. \\
  The results presented here were calculated on a lattice of size
  $64 \times 64$ lattice sites and with an energy
resolution of $4096$
  points in the interval $[-30t, 30t]$.
\end{thebibliography}
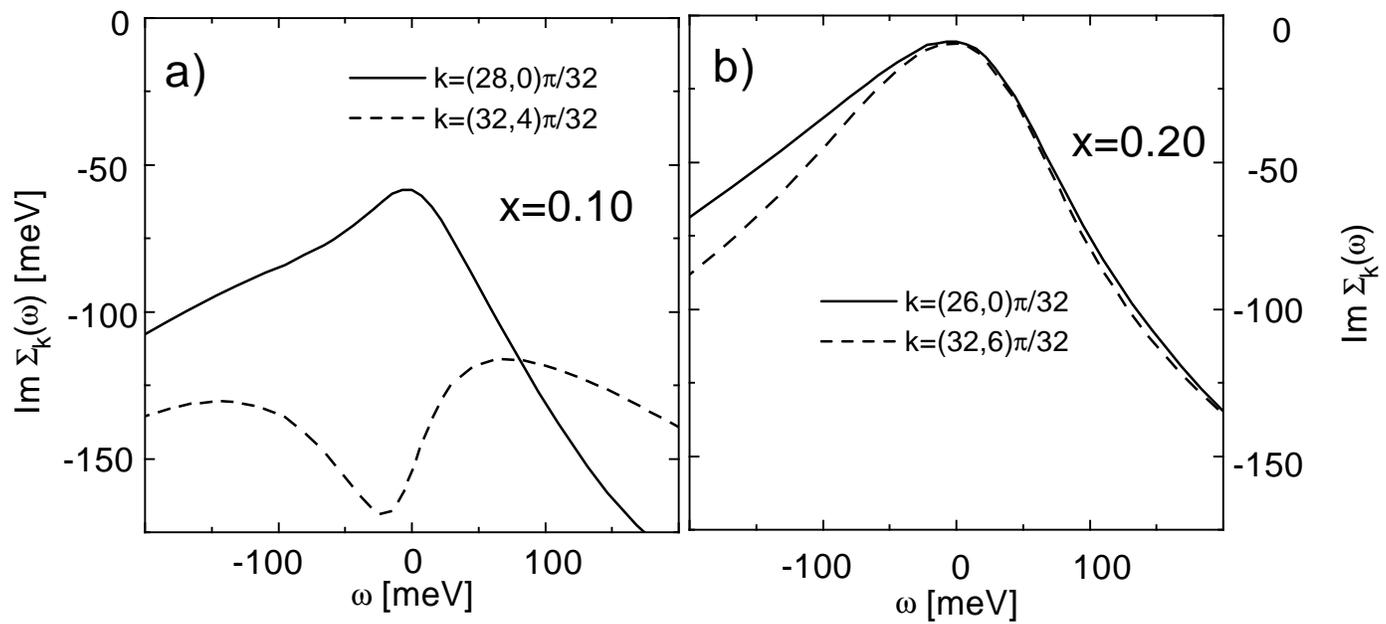
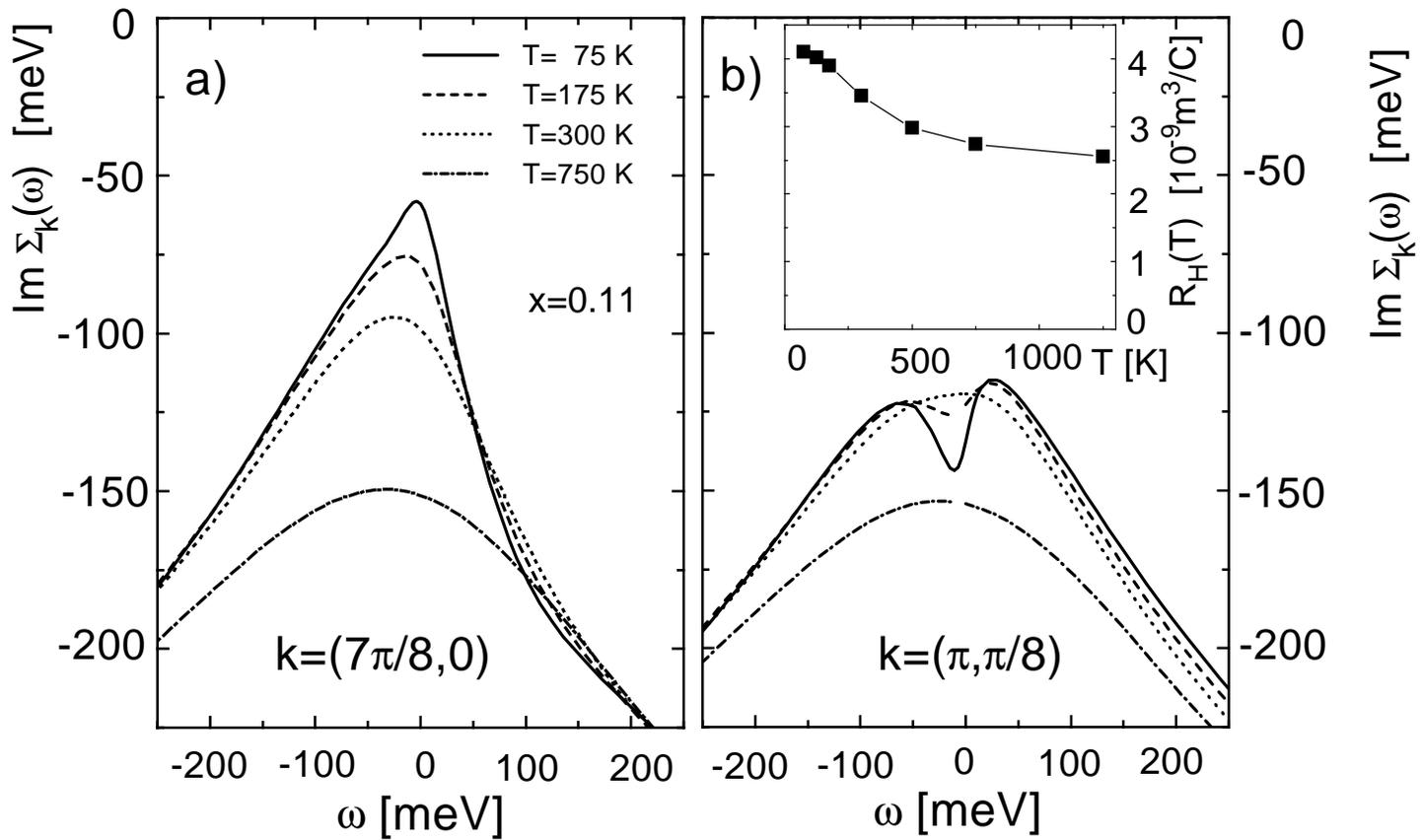
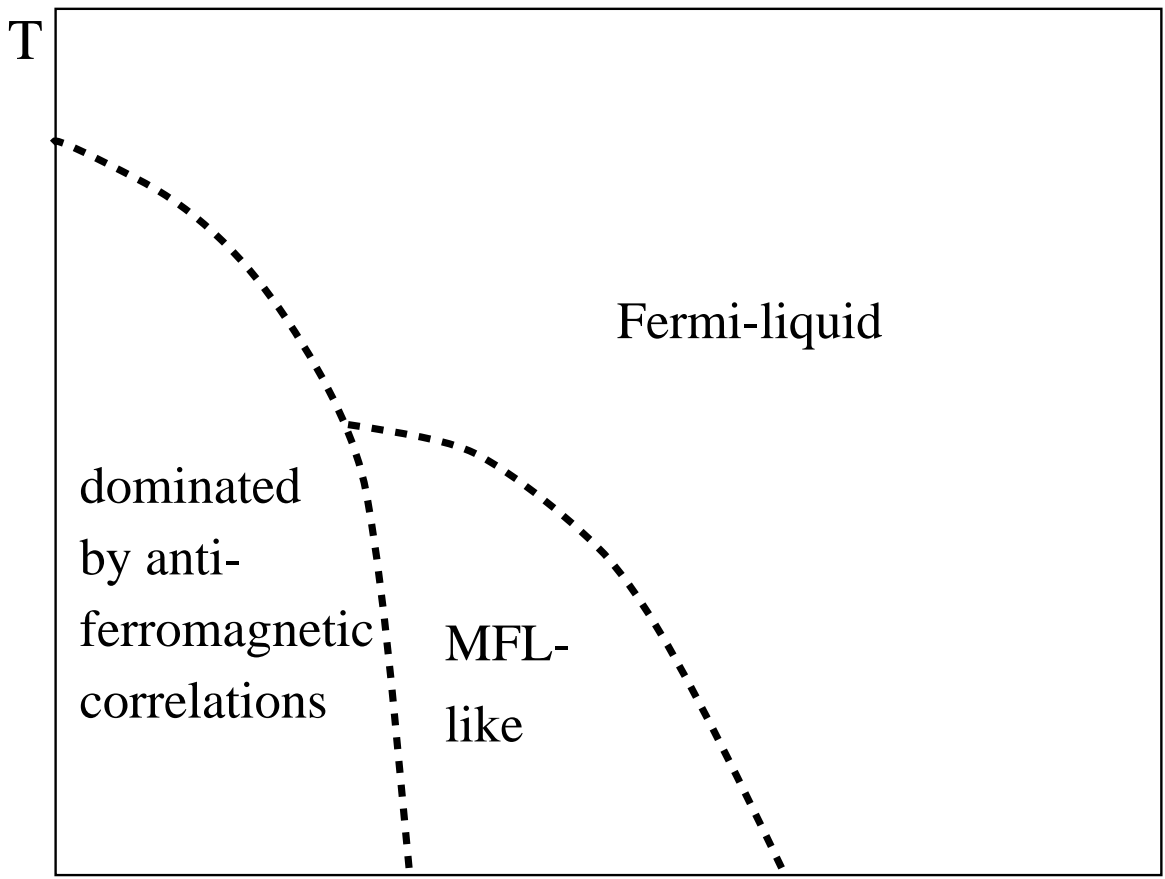
\end{document}